\documentclass[twocolumn,preprintnumbers,notitlepage,superscriptaddress,amsmath,10pt,aps,prl]{revtex4-1}
\usepackage{graphicx}
\usepackage{color}
\usepackage{hyperref}
\usepackage{subfigure}
\usepackage{graphicx}
\usepackage{type1cm}
\usepackage{eso-pic}
\usepackage{color}
\usepackage{amsmath}
\usepackage{amssymb}
\usepackage{textcomp}

\begin{document}

\title{Optimal simultaneous measurements of incompatible observables of a single photon}
%\title{Quantum-limited joint measurements of incompatible observables of a single photon}

%\affiliation{Centre for Quantum Photonics, H. H. Wills Physics Laboratory and Department of Electrical and Electronic Engineering, University of Bristol, Bristol, BS8 1TL, United Kingdom}
%\affiliation{Institute for Photonics and Quantum Sciences, SUPA, Heriot-Watt University, Edinburgh EH14 4AS, United Kingdom}

\author{Adetunmise C. Dada}
\email{adetunmise.dada@bristol.ac.uk}
\affiliation{Centre for Quantum Photonics, H. H. Wills Physics Laboratory and Department of Electrical and Electronic Engineering, University of Bristol, Bristol, BS8 1TL, United Kingdom}

\author{Will McCutcheon}
\affiliation{Centre for Quantum Photonics, H. H. Wills Physics Laboratory and Department of Electrical and Electronic Engineering, University of Bristol, Bristol, BS8 1TL, United Kingdom}

\author{Erika Andersson}
\affiliation{Institute for Photonics and Quantum Sciences, SUPA, Heriot-Watt University, Edinburgh EH14 4AS, United Kingdom}
\author{Jonathan Crickmore}
\affiliation{Institute for Photonics and Quantum Sciences, SUPA, Heriot-Watt University, Edinburgh EH14 4AS, United Kingdom}
\author{Ittoop Puthoor}
\affiliation{Institute for Photonics and Quantum Sciences, SUPA, Heriot-Watt University, Edinburgh EH14 4AS, United Kingdom}
\author{Brian D. Gerardot}
\affiliation{Institute for Photonics and Quantum Sciences, SUPA, Heriot-Watt University, Edinburgh EH14 4AS, United Kingdom}

\author{Alex McMillan}
\affiliation{Centre for Quantum Photonics, H. H. Wills Physics Laboratory and Department of Electrical and Electronic Engineering, University of Bristol, Bristol, BS8 1TL, United Kingdom}

\author{John Rarity}
\affiliation{Centre for Quantum Photonics, H. H. Wills Physics Laboratory and Department of Electrical and Electronic Engineering, University of Bristol, Bristol, BS8 1TL, United Kingdom}

\author{Ruth Oulton}
\affiliation{Centre for Quantum Photonics, H. H. Wills Physics Laboratory and Department of Electrical and Electronic Engineering, University of Bristol, Bristol, BS8 1TL, United Kingdom}

%\dates{Compiled \today}

%\ociscodes{OCIS codes: (270.5585) Quantum information and processing, (120.0120) Measurement, (270.0270) Quantum optics, (060.5565) Quantum communications, (190.4380) Four-wave mixing, (190.4370) Nonlinear optics, fibers}
%\doi{\url{http://dx.doi.org/10.1364/optica.99.999999}}

\begin{abstract}
{ Joint or simultaneous measurements of non-commuting quantum observables are possible at the cost of increased unsharpness or measurement uncertainty. Many different criteria exist for defining what an ``optimal" joint measurement is, with corresponding different tradeoff relations for the measurements. Understanding the limitations of such measurements is of fundamental interest and relevant for quantum technology. Here, we experimentally test a tradeoff relation for the \emph{sharpness} of qubit measurements, a relation which refers directly to the form of the measurement operators, rather than to errors in estimates. We perform the first optical implementation of the simplest possible optimal joint measurement, requiring less quantum resources than have previously often been employed. Using a heralded single-photon source, we demonstrate quantum-limited performance of the scheme on single quanta.}\\
\end{abstract}

%\setboolean{displaycopyright}{true}

\maketitle
%\thispagestyle{fancy}
%\ifthenelse{\boolean{shortarticle}}{\abscontent}{}

\section{Introduction}

%%%%%%%%%-------------FIGURES
There are several types of uncertainty relations in quantum mechanics. To start with, if two non-commuting observables are each measured separately, `` sharply"  (each observable measured on an ensemble of identically prepared quantum systems), then the product of their variances is bounded from below by uncertainty relations~\cite{Heisenberg1927anschaulichen,robertson1929uncertainty,EScrhodinger1930About}. %, deutsch1983uncertainty}. 
%{\bf Check that the Heisenberg paper also mentions this sort of uncertainty relation -- Erika thinks it does, but is not sure.} 
% The Deutsch paper is about an entropic uncertainty relation, not for the product of the variances - or about the variances at all, so if we want to cite this, we need a separate sentence. OK.
In addition, measurements generally disturb a measured quantum state. This leads to further limitations on how well two observables can be measured {\em jointly} on the {\em same} quantum system. Different criteria for exactly what is to be optimised lead to different uncertainty or tradeoff relations for joint measurements, see e.g.~\cite{Heisenberg1927anschaulichen, arthurskelly1965, arthursgoodman1988, stenholm1992simultaneous, hall2004prior, ozawa2003universally, busch2013proof, Busch:1986kxba, Busch:2001ue, busch2014Heisenberg}. 
%{\bf This should be a suitable selection of joint measurement uncertainty relation papers (theory, or where such relations are introduced). There are LOTS more and people are arguing quite a lot.}

Uncertainty relations apply to measurements of any non-commuting observables, such as position and momentum, and spin-1/2 (qubit) observables. Aside from their fundamental interest, uncertainty relations are relevant for quantum technology, including for quantum state estimation and quantum metrology. For example, they limit how much we can learn about different properties of quantum systems, and are related to why one can bound the information held by an eavesdropper in quantum key distribution. 

In this paper, we present the realisation of a tradeoff relation for joint measurements of a spin-1/2 system, given in~\cite{Busch:1986kxba,Busch:2001ue}. Our realization uses the polarization of heralded single photons.
Several experimental tests of different kinds of uncertainty relations for joint measurements have been reported, see for example~\cite{Erhart2012,Weston2013, Sulyok2013, rozema2012violation,Ringbauer:2014gk,Kaneda:2014emba, xiong2017optimal}.
%{\bf Continuous simultaneous measurement~\cite{hacohen2016quantum} --  relevant or not? Does not test uncertainty relations.}
Many of these realisations have used weak measurements, even if any optimal quantum measurement will necessarily be described by a specific generalised quantum measurement (POM or POVM) which always can be realized in a single shot, with no need to resort to the framework of weak measurements or postselection.  Joint measurements can also be accomplished through quantum cloning~\cite{Brougham:2006kz,thekkadath2017determining}. This usually requires   entangling operations, thereby imposing practical limitations, e.g., for  photonic quantum technologies where deterministic entangling gates are lacking.

One might also expect that in order to realise a joint measurement of two non-commuting observables, it would be necessary to couple the quantum system to be measured to an ancillary system. 
For two qubit observables, however, it turns out that this is not necessary, and that an optimal measurement can be implemented by probabilistically selecting to perform one or the other of two projective measurements~\cite{Andersson:2005kvbacadaea}. Such a setup was also suggested for measurement along two orthogonal spin directions in \cite{busch1987some,Barnett1997}. This leads to the simplest possible realisation of an optimal joint measurement, requiring no entangling operations, and is therefore the technique we employ here.

An early example of a tradeoff relation for joint measurements was given in~\cite{Busch:1986kxba,Busch:2001ue}. This relation holds for measurements on spin-1/2 systems. A related relation~\cite{busch2014Heisenberg} has been experimentally realized on a single trapped ion~\cite{xiong2017optimal}. Here, we aim to test the original relation given in~\cite{Busch:1986kxba,Busch:2001ue} using single photons.

\section{Theoretical framework}
{\bf The sharpness tradeoff relation.} 
In~\cite{Busch:1986kxba,Busch:2001ue}, the joint measurement is assumed to have marginal measurement operators of the form
\begin{equation}
\label{eq:margoper}
\Pi^a_\pm=%\Pi^{ab}_{++}+\Pi^{ab}_{+-}=
{1\over 2}(\mathbf{\hat 1}\pm \alpha {\bf a}\cdot\hat{\sigma}), ~~~
\Pi^b_\pm=%\Pi^{ab}_{++}+\Pi^{ab}_{+-}=
{1\over 2}(\mathbf{\hat 1}\pm \beta {\bf b}\cdot\hat{\sigma}), 
%\Pi^a_-&=&%\Pi^{ab}_{-+}+\Pi^{ab}_{--}=
%{1\over 2}(\mathbf{1}-\bf \alpha a\cdot\hat{\sigma})
\end{equation}
for the jointly measured spin-1/2 observables $\hat A = {\bf a}\cdot\hat\sigma $ and $\hat B = {\bf b}\cdot\hat\sigma$, where $\bf a$ and $\bf b$ are Bloch vectors, each corresponding to a qubit (e.g., polarization) state. 
It always holds that $0\le \alpha, \beta\le 1$. If $\alpha=1$, then 
$\Pi^a_\pm $ correspond to a projective measurement of \mbox{${\bf a}\cdot\hat\sigma$}, while $\alpha=0$ corresponds to a random guess, and similarly for $\Pi^b_\pm$. The coefficients $\alpha$ and $\beta$ can be referred to as the {\em sharpnesses} of the measurements of ${\bf a}\cdot\hat\sigma$ and ${\bf b}\cdot\hat\sigma$, and the closer they are to 1, the sharper the measurements.

The tradeoff relation for $\alpha$ and $\beta$ given in \cite{Busch:1986kxba,Busch:2001ue} is
\begin{equation}
\label{alphabound}
 |\alpha {\bf a} + \beta {\bf b}| + |\alpha {\bf a} - \beta {\bf b}|\leq 2,
\end{equation}
which can be rewritten as~\cite{Andersson:2005kvbacadaea}
\begin{equation}
\label{eq:alphaunc}
\Delta_\alpha^2\Delta_\beta^2 \equiv {(1-\alpha^2)(1-\beta^2)\over {\alpha^2\beta^2}}\geq\sin^2(2\theta),
\end{equation}
where $2 \theta$ is the angle between $\bf a$ and $\bf b$, and $\theta$ would be the angle between the equivalent polarization- or qubit-state vectors.
The bound in \eqref{alphabound} and \eqref{eq:alphaunc} is tight, in the sense that a joint quantum measurement with marginal measurement operators given by \eqref{eq:margoper}, for any $\alpha$ and $\beta$ saturating the bound, can always be realised.

Note that the bound %in \eqref{alphabound} or \eqref{eq:alphaunc} 
does not depend on the measured state, nor on what values we assign to the measurement outcomes. In this sense, the bound in \eqref{alphabound} and \eqref{eq:alphaunc} can be said to be more ``fundamental" than relations which depend on what values are assigned for measurement outcomes, which is the case for typical error-disturbance relations. In return, we assume %are of course assuming 
that ${\bf a}\cdot\hat\sigma$ and ${\bf b}\cdot\hat\sigma$ are jointly measured using measurement operators of the form in \eqref{eq:margoper}. More generally, however, measurement operators for a joint measurement of ${\bf a}\cdot\sigma$ and ${\bf b}\cdot\sigma$ do not have to be of the form in \eqref{eq:margoper}~\cite{hall2004prior, busch2014Heisenberg}, in which case the bound in \eqref{alphabound} and \eqref{eq:alphaunc} also retains its relevance. In fact, any dichotomic spin-1/2 observable will have measurement operators of the form
\begin{equation}
\label{eq:margopergen}
\Pi_\pm={\gamma_\pm\over 2}\mathbf{\hat 1}\pm{{\gamma_k\over 2} \bf k}\cdot\hat{\sigma},
\end{equation}
where $\gamma_+, \gamma_- \ge \gamma_k$ (since measurement operators must be positive) and $\gamma_+ + \gamma_- = 1$ (since $\Pi_++\Pi_-=\hat {\mathbf 1}$ must hold). Thus, the marginal measurement operators of a joint measurement must more generally be of this form. It turns out that if we choose $\gamma_+=\gamma_- = 1/2$, then the measurement can be made sharper, in the sense that $\gamma_k$ in the joint measurement is as large as possible, while keeping the direction $\bf k$ the same. Therefore, no matter what the measurement results are used to estimate, measurement operators of the form in \eqref{eq:margoper} can be said to be optimal. In this sense, \eqref{alphabound} and \eqref{eq:alphaunc} retain their relevance even more generally.
For example, in~\cite{busch2014Heisenberg}, one essentially ends up with measurement operators of the form \eqref{eq:margoper}, but for two ``new" spin directions other than $\bf a$ and $\bf b$, and hence one obtains a relation of the form in \eqref{alphabound} and \eqref{eq:alphaunc}, just for these two ``new" spin directions.

     \begin{figure}[t!] 
\centering{\includegraphics[width=0.7\linewidth]{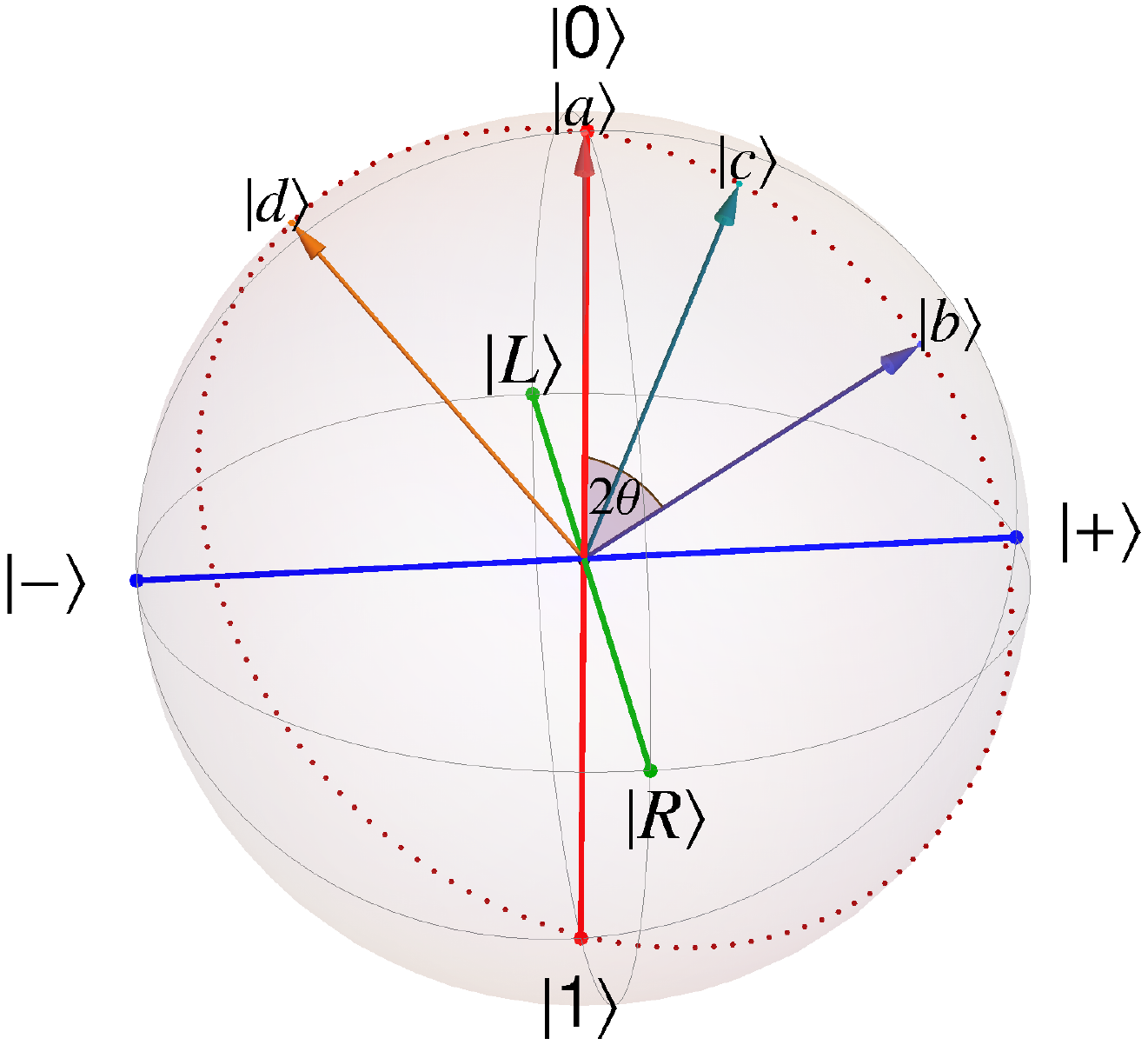}} 
\caption{{\bf Joint measurement of incompatible observables  ${\hat{A}={\bf a} \cdot\hat\sigma}$ and ${\hat{B}={\bf b}\cdot\hat\sigma}$}. The  joint measurement is implemented by doing a projective measurement \emph{either} of ${\bf c}\cdot\hat\sigma$ {\em or} of ${\bf d}\cdot\hat\sigma$, with probabilities $p$ and $1-p$ respectively, where ${\bf c}$ and ${\bf d}$ will lie in the plane defined by $\bf a$ and $\bf b$. The measurements correspond to projective measurements of photon polarization in appropriate bases. Here ${\bf a}, {\bf b}, {\bf c}, {\bf d}$ are Bloch vectors, denoted in the figure using ket representations of the corrrepsonding polarization states $|{ a}\rangle$, $|{ b}\rangle$,  $|{ c}\rangle$, $|{ d}\rangle$; each Bloch vector of unit length corresponds to a pure state. 
   } \label{fig1:exptsetupa} \end{figure}

We can also connect \eqref{eq:alphaunc} with uncertainty relations. The total uncertainties in the joint measurement, denoted $\Delta A_j$ and $\Delta B_j$, arise from two sources: the ``intrinsic uncertainties" $\Delta A$ and $\Delta B$ in the quantum observables when they are measured sharply (measured separately, not jointly) on some quantum state, and ``extra" uncertainty coming the fact that they are measured jointly.
If we assume that the measurement results both for the sharp and the joint measurements are said to be  $\pm 1$, then the variance in the joint measurement of ${\bf a}\cdot\hat\sigma$, scaled with $\alpha^{-2}$, can be written
\begin{equation}
\Delta^2 A_j /\alpha^2= (1-\alpha^2\langle\hat{A}\rangle^2)/\alpha^2 = 
(1-\alpha^2)/\alpha^2 + 1-\langle\hat{A}\rangle^2,
\end{equation}
and similarly for $\Delta^2 B_j$\footnote{Equivalently, we can assume that the measurement results are $\pm 1/\alpha$}.
Here $1-\langle{\hat{A}}\rangle^2=\Delta^2A$ is the variance of $\hat{A}$, when measured separately and sharply, and similarly for $\hat{B}$. The quantities $(1-\alpha^2)/\alpha^2\equiv \Delta_\alpha^2$ and $(1-\beta^2)/\beta^2\equiv \Delta_\beta^2$ are seen to be contributions coming from the fact that the measurement is a joint measurement. A lower bound on their product is given by (\ref{eq:alphaunc}), which can now be interpreted as an uncertainty relation giving a lower bound on the uncertainty associated purely with the fact that quantum observables $\hat{A}$ and $\hat{B}$ are measured {\it jointly}.

{\bf Optimal joint-measurement scheme.} It turns out that an optimal joint measurement along spin directions ${\bf a}$ and ${\bf b}$ can be realized by doing a projective measurement \emph{either} along $\bf{c}$ {\em or} along $\bf{d}$ %(i.e., in polarization basis $\{|c\rangle,|c^\perp\rangle\}$, or $\{|d\rangle,|d^\perp\rangle\}$) 
 with probability $p$ or $1-p$ respectively, as illustrated in Fig.~\ref{fig1:exptsetupa}~\cite{Andersson:2005kvbacadaea}.  The results of  the joint measurement are assigned %based on the measurement outcomes 
 as follows. If measurement along $\bf{c}$ is chosen, and the outcome is  $C=+1$, then the result of the joint measurement is  $A_j=+1 $ and $B_j=+1 $. If the outcome is  $C=-1$, the result of the joint measurement is  $A_j=-1 $ and $B_j=-1 $. However, if the selected measurement is along $\bf{d}$, and the outcome is $D=+1$, then result of the joint measurement is $A_j=+1 $ and $B_j=-1$, while $D=-1$ corresponds to $A_j=-1 $ and $B_j=+1 $. 
   
%It is not a priori clear that this procedure is optimal, but it turns out to be so. 
The expectation values for this joint measurement are then
\begin{align}
\overline{A_j}  = p\langle{{\bf c}\cdot\hat\sigma}\rangle +  (1-p)\langle {{\bf d}\cdot\hat{\sigma}}\rangle,\nonumber\\
\overline{B_j}  = p\langle{{\bf c}\cdot\hat{\sigma}}\rangle -  (1-p)\langle {{\bf d}\cdot\hat{\sigma}}\rangle.
\label{eq1:jmconstr}
\end{align}
On the other hand, the expectation values for a joint measurement with marginal measurement operators given by \eqref{eq:margoper} are $\overline{A_j}=\alpha\langle {\bf a}\cdot\hat\sigma\rangle$ and $\overline{B_j}=\beta\langle {\bf b}\cdot\hat\sigma\rangle$. We therefore obtain
%This POVM strategy results in the optimum joint quantum measurement if, given $\bf{a}$ and $\bf{b}$, we implement it using $p$ and measurement directions  $\bf{c}$,  $\bf{d}$ such that~\cite{Andersson:2005kvbacadaea}
\begin{align}
 {\bf c} =\frac{\left( \alpha  {\bf a} + \beta  {\bf b} \right)}{2p} {\rm ~~~and~~~}
 {\bf d} = \frac{\left( \alpha  {\bf a} - \beta  {\bf b} \right)}{2(1-p)}. 
\label{eq1:jmconstruct}
\end{align}
Since ${\bf c},~{\bf d}$ are unit vectors, it holds that
\begin{eqnarray}
p &=&|(\alpha {\bf a} + \beta {\bf b})|/2\nonumber\\
1-p &=&|(\alpha {\bf a} - \beta {\bf b})|/2.
\label{eq:pexpr}
\end{eqnarray}
Adding these two equations, we see that this measurement satisfies equality in \eqref{alphabound}, meaning that the joint measurement realized through measuring either $\hat{C}={\bf c}\cdot\hat\sigma$ or $\hat{D}={\bf d}\cdot\hat\sigma$ with probabilities $p$ and $1-p$ is indeed optimal.

\begin{figure}[t!] 
\includegraphics[width=1\linewidth]{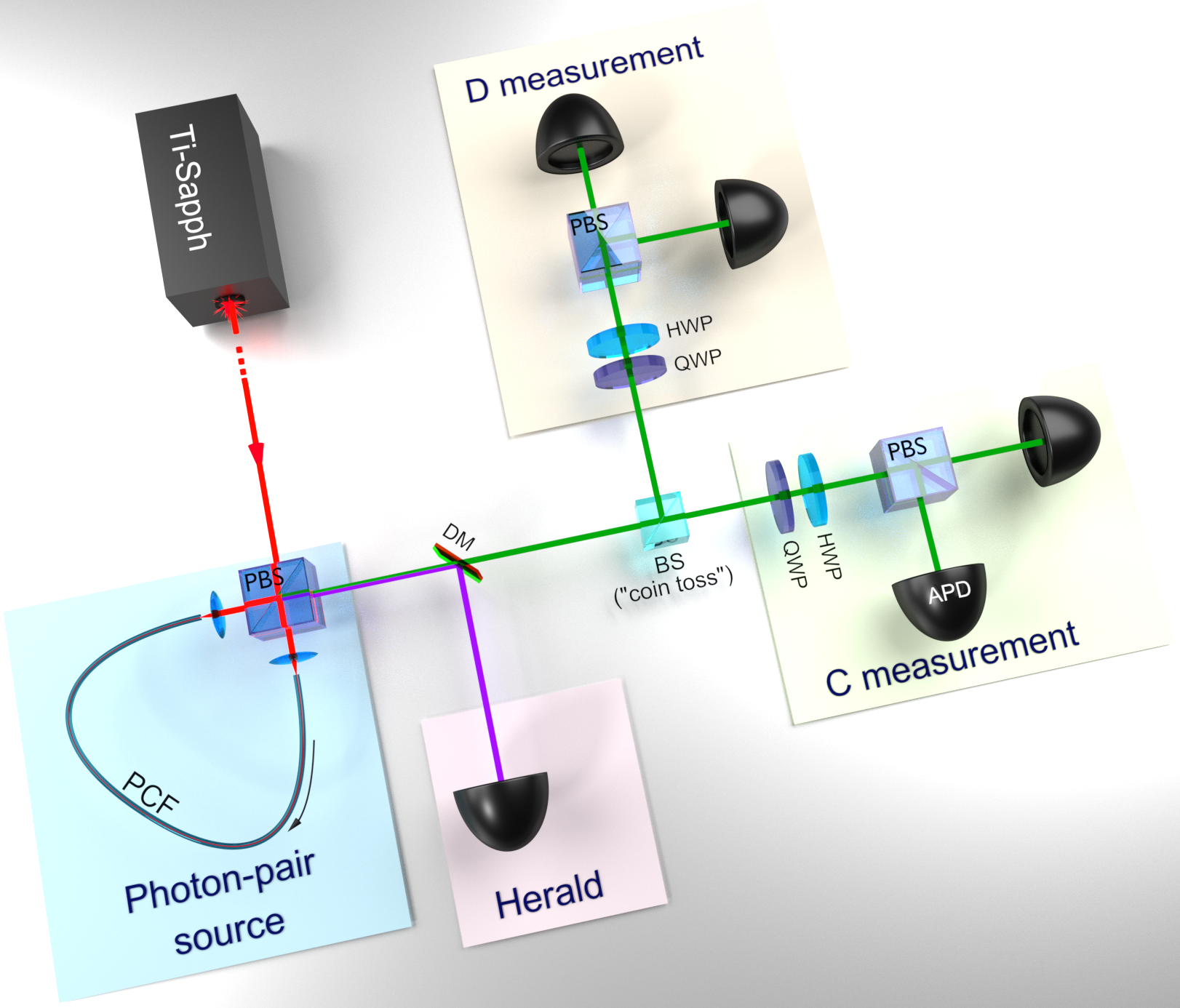} 
\caption{{ {\bf Schematic of experimental realization of a deterministic scheme for joint quantum measurements:}  
Single signal photons (623~nm) heralded by idler photons (871~nm) were generated from a four-wave mixing source in a photonic-crystal fiber pumped at 726~nm. To realize the joint measurement of observables ${{\bf a}\cdot\hat{\sigma}}$ and ${\bf b}\cdot\hat{\sigma}$, %(i.e. simultaneous measurement in bases $\{|a\rangle,|a^\perp\rangle\}$  and $\{|b\rangle,|b^\perp\rangle\}$, where  ${| a\rangle}$ and ${| b \rangle}$ are non-orthogonal state vectors), 
the signal photons, prepared in a well defined polarization state, are measured in either a polarization basis corresponding to a measurement of  ${\bf c}\cdot\hat\sigma$ or to that of ${\bf d}\cdot\hat\sigma$,
%the bases $\{|c\rangle,|c^\perp\rangle\}$ ("C measurement''), \emph{or} $\{|d\rangle,|d^\perp\rangle\}$ ("D measurement''), selected randomly 
with probabilities $p,~1-p$ respectively.  The random selection probability corresponding to the splitting ratio of the beam splitter is $p\sim 0.7$, but could also be implemented with a toss of a unbalanced classical coin.  
The source generates photon pairs cross-polarised to the pump which are filtered by both the PBS and additional wideband filters (not shown) resulting in pure horizontally polarised heralded photons entering the measurement stage.} Components: Pulsed laser (Ti-Sapph), half-wave plate (HWP), quarter-wave plate (QWP), polarizing beam splitter (PBS), non-polarizing beam splitter (BS), dichroic mirror (DM), multi-mode-fiber-coupled single-photon avalanche diode (APD). } \label{fig1:exptsetup} \end{figure}

Solving \eqref{eq:pexpr} and using equality in (\ref{eq:alphaunc}), we can express the corresponding optimal $\alpha$ and $\beta$ in terms of $\theta$ and $p$ as 
\begin{align}
 \label{eq:albeSolns}
\alpha_{\rm opt}=&\frac{(2 p-1) }{\beta_{\rm opt} \cos (2 \theta )}, \text{ where} \nonumber\\
\beta_{\rm opt} =&\pm \{\sqrt{\pm[2 (p-1) p+1]^2-(1-2 p)^2 \sec ^2(2 \theta)}+\\\nonumber
 &~~~~~~~~+2 (p-1) p+1\}^{\frac{1}{2}}.
\end{align}

\section{Experiment}

The schematic of the experimental measurement setup realizing the strategy is shown in Fig.~\ref{fig1:exptsetup}.  
We prepare the input state using a combination of waveplates (not shown) on the input arm before the beam splitter, and implement the random selection between measurement directions  $\bf{c}$,  $\bf{d}$ using a fixed, non-polarizing beam splitter with a splitting ratio corresponding to $p\sim0.7$.    
The reason for choosing $p\sim 0.7$ is that this choice allows us to investigate a range of angles between the directions $\bf a$ and $\bf b$, by varying the directions $\bf c$ and $\bf d$. In return, the maximum possible angle between $\bf a$ and $\bf b$ we can achieve is about $2\theta = 50^\circ$. %(see Supplementary Material). 
If one would want to perform a joint measurement of maximally complementary observables, this can only be achieved with $p=1/2$. Conversely, $p=1/2$ would always result in a measurement of two maximally complementary spin-1/2 observables; by varying the directions $\bf c$ and $\bf d$, one can in that case vary the relative sharpnesses of the measurements of ${\bf a}\cdot\hat\sigma$ and ${\bf b}\cdot\hat\sigma$ in the joint measurement.

To determine the optimal measurement directions ${\bf c}$ and ${\bf d}$, we solve the equations \eqref{eq:pexpr}  for $\alpha,\beta$ satisfying equality in \eqref{eq:alphaunc} for each combination of $\bf{a},\bf{b}$. For each $\bf{a},\bf{b}$, we then use these solutions $\alpha_{\rm opt}$, $ \beta_{\rm opt}$, i.e., \eqref{eq:albeSolns}, in   \eqref{eq1:jmconstruct} to get the  $\bf{c},\bf{d}$ that are subsequently used as the measurement settings. Of course, due to experimental imperfections, the actual experimental values ${\alpha}_{\rm exp},{\beta}_{\rm exp}$ determined from the measurements of $\bf{c},\bf{d}$ chosen in this way may not necessarily saturate the bound in \eqref{eq:alphaunc}. However, we are able to saturate this bound within experimental error bars from Poissonian photon counting statistics. 
Note that for this scheme it would suffice to use a classical random selection of the measurement, and this is equivalent to the toss of an unbalanced classical coin. 

\begin{figure}[ht!]
\centering{\includegraphics[width=0.8\linewidth]{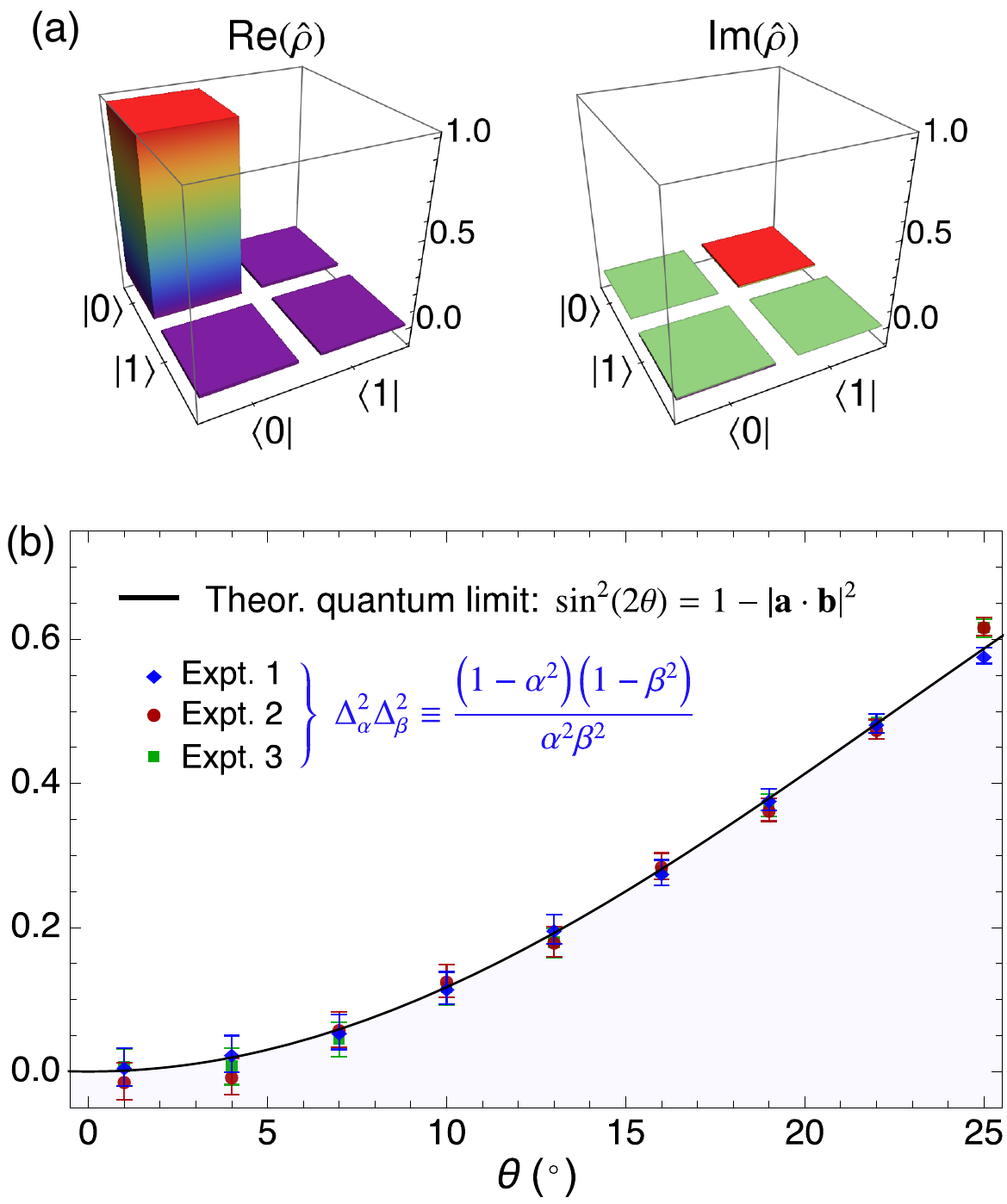}}
\caption{{{\bf Experimental results}: {\bf (a) Input state tomography.} 
Tomography has been performed in the the ``C measurement" arm of the apparatus as well as the ``D measurement" arm resulting in state reconstructions having fidelity $\mathcal{F}=99.9993(2)\%$ confirming that the measurement bases are well calibrated to  each other.
{\bf (b) Sharpness of the joint measurement.} RHS of \eqref{eq:alphaunc} as a function of $\theta$, where $\theta = \arccos(|{\bf a}\cdot{\bf b}|)/2 $, with error bars determined only by Poisson statistics of raw count rates for three sets of experiments each corresponding to ${\bf a}$, ${\bf b}$ defining a distinct plane on the Bloch sphere (see Fig.~\ref{figS1:BlochABCDstates}).  Each data point is an average of $100$ runs, each of which involved the detection of $\sim1.5\times 10^4$ heralded single photons. 
 The solid black line represents the quantum limit. Data below this line are forbidden by quantum theory.  
 } }
\label{fig3:VarVSoverlap}
\end{figure}
\begin{figure}[ht!]
\centering{\includegraphics[width=1\linewidth]{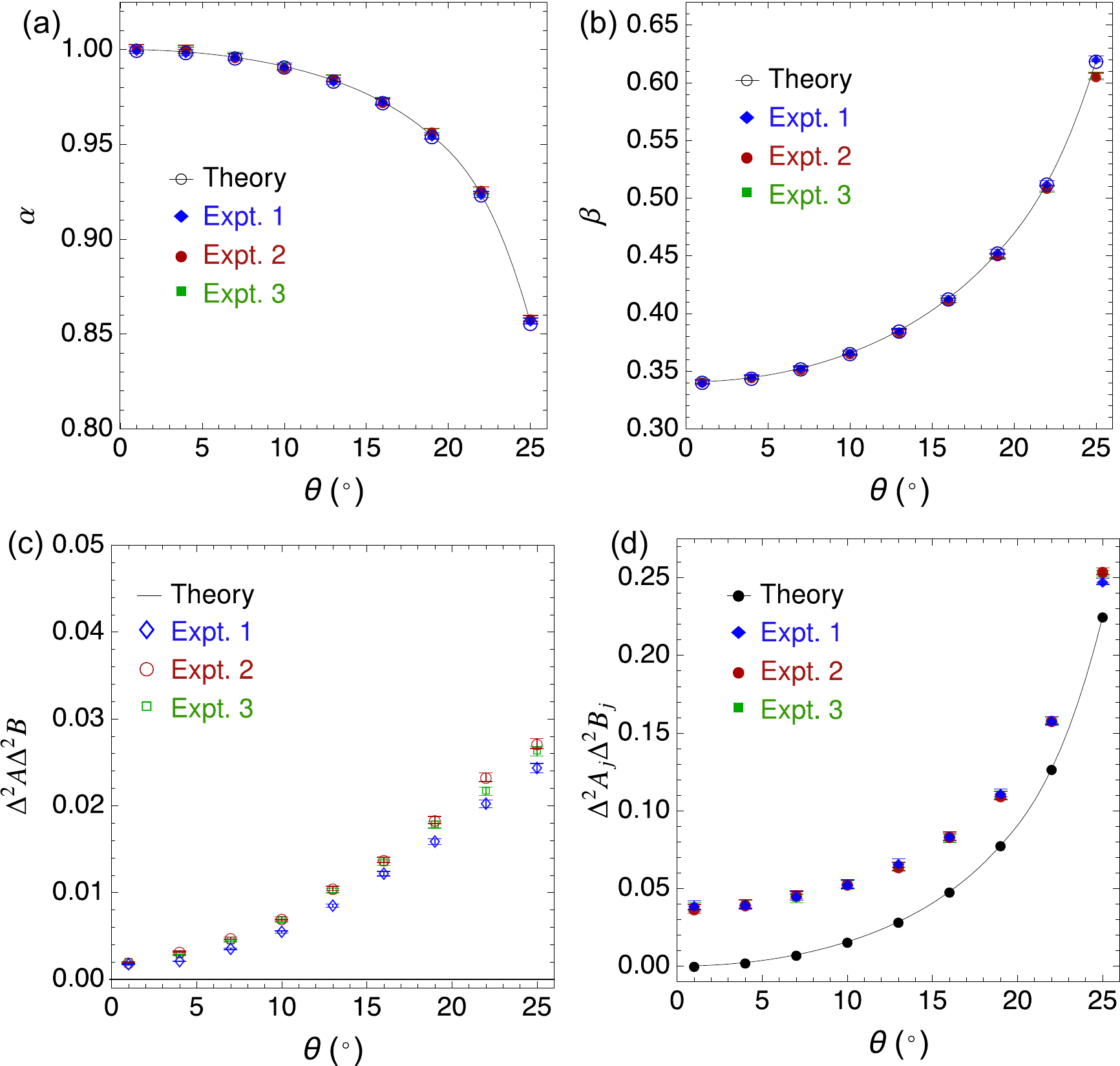}}
\caption{Experimental values of (a) $\alpha$ and (b) $\beta$ as functions of $\theta$, plotted with the corresponding theoretical values. (c) Product of experimental variances for the sharp measurements. (d) Product of experimental variances for the joint measurements and their comparison with theory. }
\label{figS4:gRowAlphasBetas}
\end{figure}

We perform three sets of experiments using pairs ${\bf a},{\bf b}$, with $\bf a$ kept constant as the ${\bf z}$ direction, and varying $\bf  b$  to traverse $\theta=1,\ldots,25$\textdegree, along a different plane on the Bloch sphere for each experiment corresponding to azimuthal angles $\phi_1=-160.7$\textdegree, %-0.893\pi$,
$\phi_2=-51.6$\textdegree, % $\phi_2=-0.287\pi$ 
$\phi_3=83.7$\textdegree,  %$\phi_3=0.465\pi$
for experiments 1, 2 and 3,  respectively. 
For our experiments, we chose as input state the eigenstate of ${\bf a}\cdot \hat{\sigma}$, denoted $|a\rangle$, which coincides with the ``$|0\rangle$" state.
 We carry out the  measurements of ${\bf c}\cdot\hat\sigma$ and ${\bf d}\cdot\hat\sigma$, 
 by measuring in the correpondng polarization bases   using appropriate  settings of the half-wave plates and quarter-wave plates and subsequent measurement in the $\{|H\rangle,|V\rangle\}$ polarization basis using a polarizing beam splitter and fiber-coupled single-photon detectors.  
Using coincidence detection with idler photons as  heralds, we are able to register any of the  four possible outcomes for each heralded photon going through the measurement circuit (see Fig.~\ref{fig1:exptsetup}).

%Figure
\begin{figure*}[ht!]
\centering{\includegraphics[width=0.8\linewidth]{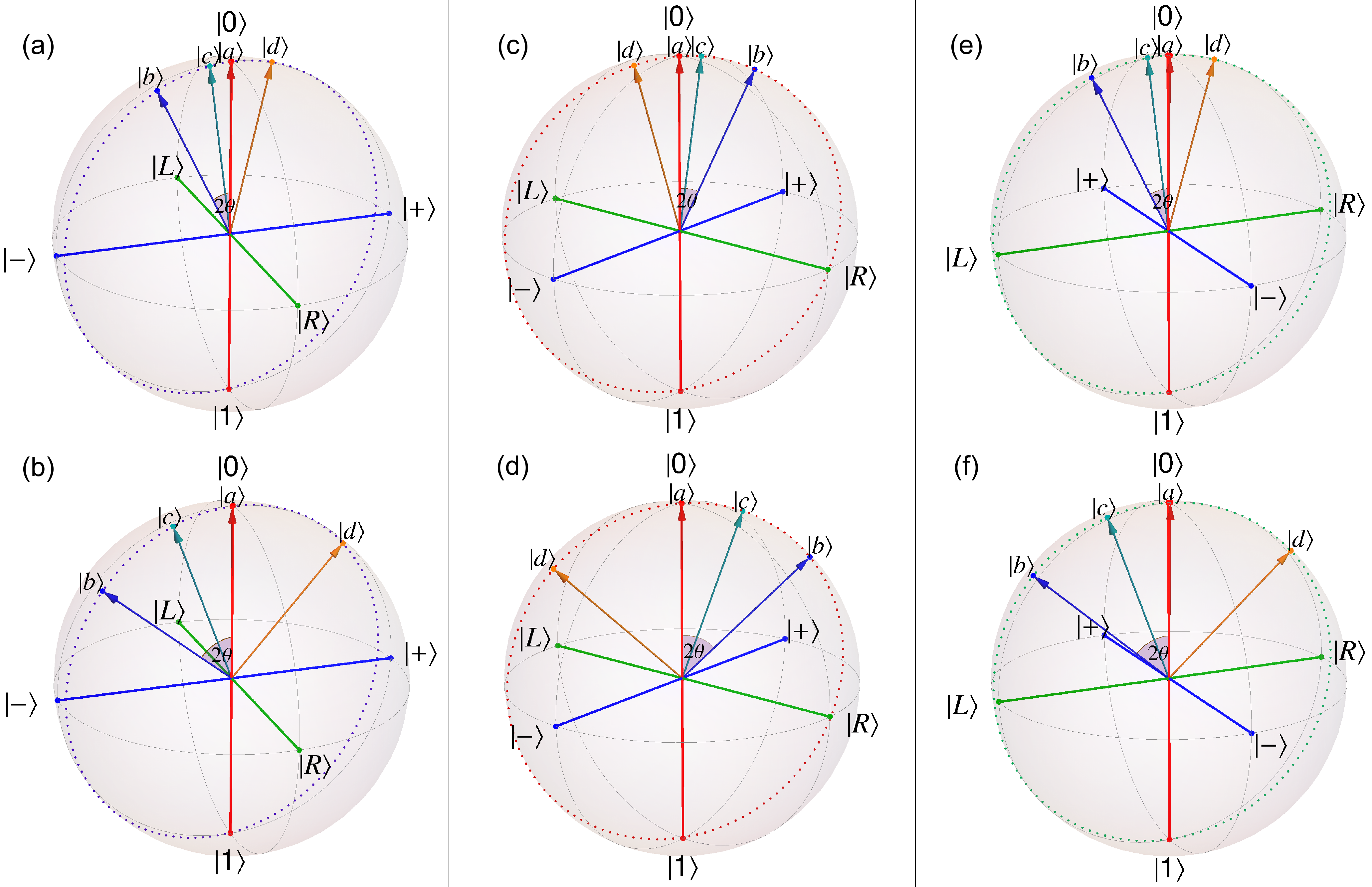}}
\caption{ {\bf Examples of pairs of incompatible observables used in the experiments.} Spin directions  $\bf{a}$,$\bf b$ (states $|a\rangle$,$|b\rangle$) defining the incompatible observables, along with $\bf{c}$, $\bf d$ (states $|c\rangle$,$|d\rangle$) for implementing their joint measurements with  (a) $\theta = 13$\textdegree~in experiment 1 (b) $\theta = 25$\textdegree~in experiment 1 (c) $\theta = 13$\textdegree~in experiment 2 (d) $\theta = 25$\textdegree~in experiment 2 (e) $\theta = 13$\textdegree~in experiment 3 (f) $\theta = 25$\textdegree~in experiment 3.  
For the experiments, 
$\bf a$  is kept constant and $\bf b$  is varied such that $\theta \equiv \arccos(|{\bf a}\cdot{\bf b}|)/2 = 1,4,7,\ldots,25$\textdegree, for each of the three sets of experiments, corresponding to azimuthal angles $\phi_1=-160.7$\textdegree, %-0.893\pi$,
$\phi_2=-51.6$\textdegree, % $\phi_2=-0.287\pi$ 
$\phi_3=83.7$\textdegree,  %$\phi_3=0.465\pi$
for experiments 1, 2 and 3, respectively.  
} 
\label{figS1:BlochABCDstates}
\end{figure*}
%\vspace{48pt}

%
\section{Methods}
%{\bf Heralded single-photon source.} 
We ensure a true single-photon implementation by exploiting a heralded source of single photons consisting of a microstructured  photonic crystal fibre (PCF) exploiting birefringent phase-matching~\cite{halder2009nonclassical,clark2011intrinsically} to produce  photon pairs via  spontaneous four-wave mixing (SFWM).  This source is pumped by a pulsed Ti-Sapphire laser with a repetition rate of 80~MHz. The fiber is highly birefringent ($\Delta n = 4 \times 10^{-4}$) with  phase-matching conditions leading to generation of signal-idler pairs with polarization orthogonal to that of the pump. 
In addition to this birefringence,  the waveguide contributions to the dispersion can be used to tailor the SFWM for generation of naturally narrowband, spectrally uncorrelated photons when pumped with Ti-Sapphire laser pulses at the flat region  of the phase-matching curves ($\lambda_{\rm pump} \simeq 726$~nm) where the idler photons ($\lambda_i = 871$~nm) are group-velocity matched to the pump pulse  so that they become spectrally broad ($\Delta \lambda_i = 2.2$~nm) while the signal photons ($\lambda_s = 623$~nm) are intrinsically narrow-band ($\Delta \lambda_s = 0.3$~nm). This narrowband phase-matching results in a highly separable joint-spectral amplitude for a wide range of pump bandwidths, thereby enabling the generation of single photons of high state purity.  
Although the fibre source is in a Sagnac-loop configuration allowing for generation of entangled states when the pump pulse is set to diagonal polarization and split at the polarising beam-splitter (Fig.~\ref{fig1:exptsetup}), we use horizontally polarised pump pulses so that the PCF is pumped in only one direction for use as a heralded single-photon source.

\begin{figure*}[ht!]
\includegraphics[width=1\linewidth]{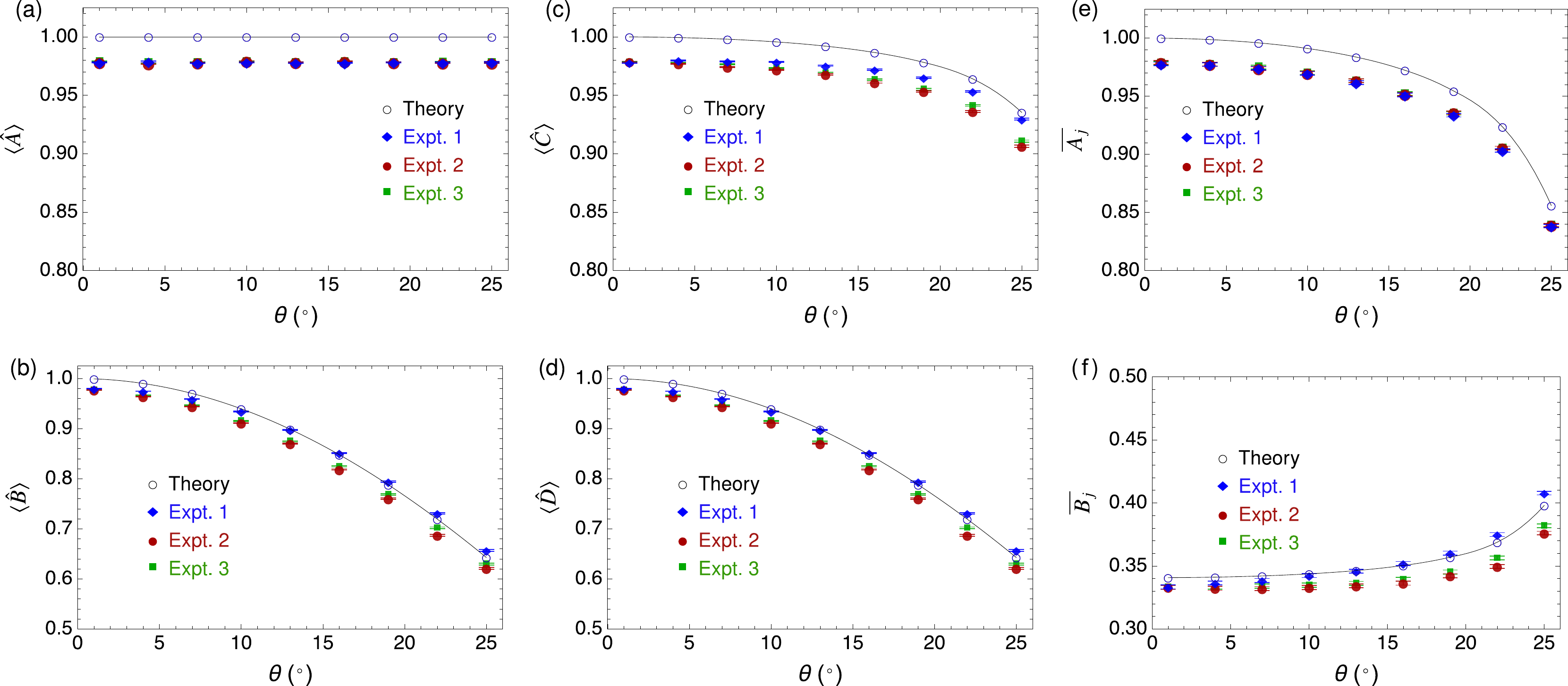}
\caption{ {\bf Expectation values for the individual ``sharp" and joint measurements}. For the three experiments, the plots show expectation values for sharp measurements of  
(a) ${\bf a}\cdot \hat{\sigma}$ and (b) ${\bf b}\cdot \hat{\sigma}$, (c) ${\bf c}\cdot\hat\sigma$ and (d) ${\bf d}\cdot\hat\sigma$, and expectation values for the  joint measurements (e) $ \overline{A_j}$ and (f) $\overline{B_j}$.  Also plotted  for comparison are the ideal theoretical predictions, which do not include the effects of experimental imperfections.}
\label{figS2:AllExpVals}
\end{figure*}

\section{Results}
Let the heralded detector count rates corresponding to ${ C}=\pm1$ and ${ D}=\pm1$ be  $\mathcal{C}_{\pm}$ and $\mathcal{D}_{\pm}$, respectively. From these, we determine the experimental expectation values as
\begin{equation}
\langle{\hat{C}}\rangle = \frac{\mathcal{C}_{+}-\mathcal{C}_{-}}{\mathcal{C}_{+}+\mathcal{C}_{-}},~~~\langle{\hat{D}}\rangle = \frac{\mathcal{D}_{+}-\mathcal{D}_{-}}{\mathcal{D}_{+}+\mathcal{D}_{-}}.
\label{eq:Cexpvalues}
\end{equation}
Using this in \eqref{eq1:jmconstr}, and using the value of $p=0.670(1)$ obtained from total count rates in the $C$, and $D$ channels,  
the experimental joint-measurement expectations values  are then  obtained directly according to \eqref{eq1:jmconstr}. 

To benchmark the performance of the implemented joint measurements, we also perform separate sharp  measurements of the incompatible observables ${\bf a}\cdot\hat{\sigma}$, ${\bf b}\cdot\hat{\sigma}$.  
 Again, if we denote the detector count rates corresponding to ${ A}=\pm1$ and ${B}=\pm1$ as  $\mathcal{A}_{\pm}$ and $\mathcal{B}_{\pm}$, respectively, the expectations values for the sharp measurements are 
\begin{equation}
\langle{{\bf a}\cdot\hat{\sigma}}\rangle = \frac{\mathcal{A}_{+}-\mathcal{A}_{-}}{\mathcal{A}_{+}+\mathcal{A}_{-}},
~~~\langle{{\bf b}\cdot\hat{\sigma}}\rangle = \frac{\mathcal{B}_{+}-\mathcal{B}_{-}}{\mathcal{B}_{+}+\mathcal{B}_{-}}.
\label{eq:Aexpvalues}
\end{equation}
This allows us to obtain experimental values of  $\alpha$, $\beta$ which directly indicates by how much the sharpnesses are worsened solely by the fact that the measurement is joint, and to evaluate the LHS of relation \eqref{eq:alphaunc}, which we plot as a function of $\theta$ in Fig.~\ref{fig3:VarVSoverlap}(b) as our main result.

Fig.~\ref{figS4:gRowAlphasBetas} shows $\alpha,~\beta$ and the product of total variances for separate (sharp) and joint measurements.  The ideal theoretical product of total ``intrinsic" variances for sharp measurements is zero, as indicated by the solid black line in Fig.~\ref{figS4:gRowAlphasBetas}(c),  since the measured state is an eigenstate of ${\bf a}\cdot\hat\sigma$, while that for the joint measurements (determined by the incompatibility of the jointly measured observables, and parametrised by $\theta$) is plotted with the black filled circles in  Fig.~\ref{figS4:gRowAlphasBetas}(d). Fig.~\ref{figS1:BlochABCDstates} shows examples of pairs of spin directions ${\bf a},{\bf b},{\bf c},{\bf d}$ used in the sets of experiments.  
Also, shown in Fig.~\ref{figS2:AllExpVals} are the expectation values for the individual ``sharp" measurements of ${\bf a}\cdot\hat{\sigma}$, ${\bf b}\cdot\hat{\sigma}$, and expectation values resulting from the implemented joint-measurement strategy. %, showing reasonable agreement with theory. 

 \section{Discussion}
The non-ideal experimental values of $\Delta^2A \Delta^2B$, rather than the joint measurement strategy, is accountable for the deviation of the $\Delta^2A_j \Delta^2B_j$ from the ideal value  because, as seen in Fig.~\ref{fig3:VarVSoverlap}, the contribution purely due to the jointedness of the measurement is at the quantum limit.

We see that, even without making any other corrections for experimental imperfections,  our results verge on the quantum mechanical limit of how much variances must increase due to performing the quantum measurements jointly. This is thanks to the simplicity of the scheme, the brightness of the heralded single-photon source which reduced the effect of Poissonian noise, and the precise calibration of the two measurement setups with each other with a fidelity of $99.9993(2)\%$.    
We emphasise that  
the quantity $\Delta^2_\alpha\Delta^2_\beta$ is extremely sensitive to experimental error and has no upper bound. 

In conclusion, 
we have demonstrated an optimal joint measurement scheme that does not use filtering or postselection, nor does it need entangling interactions with an ancilla.  
A true joint measurement of two observables performed on a single qubit or spin-$\tfrac{1}{2}$ system should have four possible outcomes for each qubit measured, as is now demonstrated here. 
 The non-requirement for 2-dimensional arrays of (single)-photon detectors, as often used in weak-measurement-based setups, also makes our scheme easier to implement. 
 The implemented scheme can easily be applied to other qubit degrees of freedom and other two-level quantum systems, since standard projective measurements and flips of unbalanced classical ``coins'"are generally easy to implement. It would be of interest to extend this scheme to joint measurement of incompatible observables of higher-dimensional systems (qudits), or of multiple incompatible observables.  
This work  demonstrates how to fully implement joint measurements of non-commuting spin-1/2 (or qubit) observables in an optimal way, and with less quantum resources than often employed previously. 

{\bf Acknowledgements:} This  work  was  funded by  the  Future  Emerging  Technologies (FET)-Open  FP7-284743  project  Spin  Photon  Angular  Momentum  Transfer  for  Quantum  Enabled  Technologies  (SPANGL4Q) and the Engineering and Physical Sciences Research
Council (EPSRC) (EP/L024020/1, EP/M024156/1,   EP/N003381/1 and EP/M024458/1).

%===========================================================================================
%\bibliographystyle{apsrev4tun}
%\bibliographystyle{osajnl}
%\bibliography{ReferencesJMv2}
%===========================================================================================


%===========================================================================================

\end{document}